# What do we mean by "data"?
# A proposed classification of data types
# in the arts and humanities

Bianca Gualandi[1], Luca Pareschi[2], and Silvio Peroni[3,4]

[1] University of Bologna, Bologna, Italy

[2] University of Rome Tor Vergata, Rome, Italy

[3] Research Centre for Open Scholarly Metadata, Department of Classical Philology and Italian Studies, University of Bologna, Bologna, Italy

[4] Digital Humanities Advanced Research Centre (/DH.arc), Department of Classical Philology and Italian Studies, University of Bologna, Bologna, Italy

## Author note

Bianca Gualandi, https://orcid.org/0000-0001-8202-8493 Email: bianca.gualandi4@unibo.it
Luca Pareschi, https://orcid.org/0000-0002-4402-9329 Email: luca.pareschi@uniroma2.it
Silvio Peroni, https://orcid.org/0000-0003-0530-4305 Email: silvio.peroni@unibo.it

# Abstract

**Purpose:** This article describes the interviews we conducted in late 2021 with 19 researchers at the Department of Classical Philology and Italian Studies at the University of Bologna. The main purpose was to shed light on the definition of the word "data" in the humanities domain, as far as FAIR data management practices are concerned, and on what researchers think of the term. **Methodology:** We invited one researcher for each of the official disciplinary areas represented within the department and all 19 accepted to participate in the study. Participants were then divided into 5 main research areas: philology and literary criticism, language and linguistics, history of art, computer science, archival studies. The interviews were transcribed and analysed using a grounded theory approach. **Findings:** A list of 13 research data types has been compiled thanks to the information collected from participants. The term "data" does not emerge as especially problematic, although a good deal of confusion remains. Looking at current research management practices, methodologies and teamwork appear more central than previously reported. **Originality:** Our findings confirm that "data" within the FAIR framework should include all types of input and outputs humanities research work with, including publications. Also, the participants to this study appear ready for a discussion around making their research data FAIR: they do not find the terminology particularly problematic, while they rely on precise and recognised methodologies, as well as on sharing and collaboration with colleagues.

**Type:** Article
**Category:** Research paper

# Introduction

The start of the discussion around widening public access to research can be traced back to the Budapest Open Access Initiative (BOAI) that, in a 2002 declaration, defined for the first time the term open access (Chan *et al.*, 2002). Other declarations followed suit (e.g., Brown *et al.*, 2003; Max Planck Society, 2003), broadening the definition to also include:

> research results, raw data and metadata, source materials, digital representations of pictorial and graphical materials and scholarly multimedia material. (Max Planck Society, 2003)

With focus shifting away from publication and towards all research materials, the terms open data and open science were born.

The European Union's Research & Innovation policy has embraced these concepts as pillars of the ERA, the European Research Area (European Commission, 2022). The new ERA policy agenda for the period 2022-2024 sets out a number of actions to "enable open science" and develop a "web of FAIR data and services for science in Europe" (European Commission, 2022; EOSC, 2022).

FAIR is an acronym for Findable, Accessible, Interoperable and Re-usable (Wilkinson *et al.*, 2016) and indicates a set of data management practices centred on machine actionability. In this scenario, the exact meaning of the word data is increasingly being discussed within the scholarly community. The use of this term, and the application of FAIR principles, within the art and humanities domain is not without problems. As pointed out, among others, by Tóth-Czifra:

> […] by applying the FAIR data guiding principles to arts and humanities data curation workflows, it will be uncovered that contrary to their general scope and deliberately domain-independent nature, they have been implicitly designed according to underlying assumptions about how knowledge creation operates and communicates. (Tóth-Czifra, 2019, p. 3)

In the same instance, the author calls for surveys "to assess whether and to what extent the term data is still a dirty word" (Tóth-Czifra, 2019, p. 15), an idea first introduced by Hofelich Mohr and colleagues in their article *When data is a dirty word: a survey to understand data management needs across diverse research disciplines* (Hofelich Mohr *et al.*, 2015).

The present work is a contribution towards this goal, and towards looking for a way forward for FAIR data principles within the arts and humanities without relying on assumptions drawn from other disciplines. This study addresses the following research questions:

1. How do we define "data" in the arts and humanities?

2. What do these researchers think of the word "data" and how do they associate "data" with research materials used in their discipline?
3. What are their attitudes towards open science?
4. What are their current data management practices?

Please note that the present study falls short of proposing solutions for the implementation of FAIR data principles in the humanities. Rather, it concentrates primarily on the definition of the term "data" and on the way it is used in humanities research communities and, secondarily, on surveying some of the current data management practices.

# Literature review

## How do we define "data" in the humanities? An open problem

The European Federation of Academies of Sciences and Humanities [1] (ALLEA) Working Group on E-Humanities, in their report *Sustainable and FAIR Data Sharing in the Humanities*, state that the definition of data encompasses all inputs and outputs of research that are not publications (Harrower *et al.*, 2020, p. 6). It does recognise, however, that both texts and documents are data (Harrower *et al.*, 2020, pp. 8 and 14).

The CO-OPERAS Implementation Network (IN), which is part of GO FAIR [2], aims at helping social sciences and humanities (SSH) communities:

> build a bridge between SSH data and the EOSC, widening the concept of "research data" to include all of the types of digital research output linked to scholarly communication that are part of the research process. (OPERAS, 2022)

In 2019-2020, they organised 5 workshops in Italy, Portugal, Germany, France, and Belgium, gathering more than 70 researchers from 32 different disciplines within SSH. They concluded that there is no consensus on what data are in these disciplines and that any definition needs to "consider the process of data production, going as deep as possible in research practices" and be "linked to the workflow of knowledge creation" (CO-OPERAS and Giglia, 2020, p. 1).

The OPERAS-P project [3], working in synergy with the CO-OPERAS IN, released a *Report on innovative approach to SSH FAIR data and publications*. The document consistently uses the expression "SSH data and publications" and confirms the importance of publications in the humanities, but its aim is ultimately to

> move from a definition of data based on data types and disciplines to the management of objects related to specific communities and aims. (Avanço, Giglia and Gingold, 2021, p. 9)

A literature review on 28 open access publications in English and French, conducted to investigate the definition of data in the SSH, indicated that "every informational data is technically data" (Avanço, Giglia and Gingold, 2021, p. 8):

> The intersection between information science and technology has changed the status and the meaning of data: anything that can be transposed into binary bits constitutes data that in turn feeds and supports the entire digital environment. At first strictly scientific as a concept and a practice, data became informational, with a wider meaning now ranging from knowledge to technology. (Avanço, Giglia and Gingold, 2021, p. 7)

Finally, the 2021 OPERAS *Report on the Future of Scholarly Communication* states that, within the FAIR framework, everything is data, or could be, with the concept of data "intended to be as universal as possible, including datasets, publications, software, etc." (Avanço, Balula *et al.*, 2021, p. 22). At the same time, it remarks that FAIR awareness is still low in SSH, and that the SSH research environment is greatly diverse and complex (Avanço, Balula *et al.*, 2021, p. 23).

It is worth mentioning at least two attempts at classifying data in this area of research. In a 2014-2015 study conducted on print and electronic dissertations in SSH submitted to the University of Lille 3 between 1987 and 2013, Prost, Malleret and Schöpfel found that 66% were accompanied by data (Prost *et al.*, 2015, p. 7). Opting for "a large and pragmatic definition" of the term (2015, p. 7), they found the following data types:

- text samples
- tables
- images (including drawings and posters)
- maps
- photographs
- statistics
- graphs (including figures, charts and visualisations)
- databases
- timelines
- audio-visual media (Prost *et al.*, 2015, pp. 12-14).

In 2019, Edmond presented this taxonomy of scholarly outputs referred, in particular, to digital humanities:

- print paradigm publications
- electronic paradigm publications
- single or collected/curated primary sources
- software
- [patents/licenses]
- [ephemera, such as exhibitions and performances] (Edmond, 2019, pp. 2-3).

Print paradigm publications are the only type to have "a clear precedent and place in the traditional flows of production, dissemination, and evaluation", while electronic paradigm ones include "everything from enhanced publications to blogs and Twitter corpora, to arguments presented in video and audio" (Edmond, 2019, p. 2). The last two categories above are enclosed in square brackets because they were initially included in the taxonomy but were discarded at a later stage (2019, pp. 2-3).

## Humanities researchers and the term "data". A problematic relationship?

While surveying researchers at Minnesota Twin City's College of Liberal Arts about their research practices, Hofelich Mohr and colleagues asked them to "self-select whether they collect, create or use 'data' or 'research materials'" (2015, p. 51). Out of 172 respondents, 54% selected "data", while 46% preferred "research materials" (Hofelich Mohr *et al.*, 2015, p. 52). A few years later, Thoegersen reached a similar conclusion while conducting a qualitative study on data management practices of humanities faculty (2018, p. 491). Most of the 9 researchers who agreed to participate, all coming from the fields of history, English, languages, and philosophy, stated "with varying levels of confidence, that they would consider most of the research materials discussed to be data" (Thoegersen, 2018, p. 498).

However, it is fair to say that the term data remains problematic in the arts and humanities. Tóth-Czifra pointed out that FAIR principles "have been implicitly designed according to underlying assumptions about how knowledge creation operates and communicates" (2019, p. 3). These assumptions are: that scholarly data or metadata is digital by nature, that scholarly data is always created and therefore owned by researchers, and that there is "a wide community-level agreement on what can be considered as scholarly data" (Tóth-Czifra, 2019, p. 3). According to Tóth-Czifra, this is not the case in the humanities where most scholarly data and metadata are held in galleries, libraries, archives, and museums, where only a fraction of them has been digitised, and where issues around their ownership remain (Tóth-Czifra, 2019, p. 3-6; Hansson and Dahlgren, 2021, p. 304). Additionally, data is a problematic concept in these disciplines and only a small community is involved in data sharing and curation (Tóth-Czifra, 2019, pp. 22-23), mostly because of a lack of incentives and of "social life of data" (2019, p. 5).

At the same time, Edmond described how "many humanists resist the term 'data' as a descriptor for their primary and secondary sources, or indeed for almost anything they produce in the course of their research". Indeed, they "already have a much richer and more nuanced vocabulary" and may find "the manner in which the term

'data' is deployed in disciplines that are primarily data-driven" problematic (Edmond, 2019, p. 6).

# Methods and materials

## Data collection

Before approaching researchers, we analysed 5 studies involving university staff in different countries and across different humanities disciplines, some of which have of course been mentioned in the previous section (Akers and Doty, 2013; Hofelich Mohr *et al.*, 2015; Prost *et al.*, 2015; Thoegersen, 2018; Tóth-Czifra and Truan, 2021).

Like others had done before us, we feared that using the term "data" might alienate at least some our interviewees. While Hofelich Mohr and colleagues (2015) had solved this problem by allowing researchers to select their favourite term, Thoegersen (2018) had opted for:

- enquiring on faculty members' research projects first,
- asking which "research materials" interviewees collected, generated, and used in their research, and how they managed them,
- introducing the term "data" only at the end, asking whether interviewees considered this term applicable to their own research.

We decided to follow this same interview structure and translated Thoegersen's questions into in Italian, with some modifications. Table I lists our questions (translated back into English) and highlights the changes made, while also making the relationship with our research questions explicit.

**Table I:** Structure of the interview, changes made, and relationship to research questions

| N | Interview questions (translated into English) | Substantial modifications made to Thoegersen (2018) | Corresponding research question |
|---|---|---|---|
| 1 | To get started, can you describe your current research project(s) for me? | none | none |
| 2 | I would like to create as complete a list as possible of the materials normally used in your discipline (*I'm thinking about, e.g., primary sources, publications, any other type of materials, in the broadest possible sense of the word*). Please list any materials you have generated or collected in the course of your research. | - added a mention to the discipline in general<br>- added a few examples of "materials", as the term is vague and potentially off-putting | RQ-1: How do we define "data" in the humanities? |
| 3 | a. Can you describe each of these materials for me? I particularly would like to know | - added a mention to format and preservation | RQ-1: How do we define "data" in the humanities? |
| | more about their format, whether physical or digital, and how/where they are held.<br>b. Are [these materials] usually accompanied by documentation? Are there any standards recognised in your discipline regarding these materials?<br>c. And, speaking of standards and good practices, are there recognised, frequently applied methodologies in your discipline? And are they somehow described or formalised? | - added a separate sub-question on documentation and standards (3b).<br>- added a separate sub-question on methodologies (3c) | RQ-4: What are humanities researchers current data management practices? |
| 4 | Who can access these materials, and do you share them (or would share them) with colleagues and other researchers? And with the public at large (e.g., online)? | none | RQ-3: What are humanities researchers' attitudes towards open science?<br>RQ-4: What are humanities researchers current data management practices? |
| 5 | Of the materials you have listed, are there any that contain confidential information or concern someone's privacy? Are there intellectual property/copyright concerns? And security concerns? | none | RQ-4: What are humanities researchers current data management practices? |
| 6 | How do you decide what research materials to keep or discard at the end of a research project? And those you keep, where do you keep them? | none | RQ-4: What are humanities researchers current data management practices? |
| 7 | Lately there is a lot of talk about "open data" and "research data". Think back at the materials we talked about; would you consider any of them "data"? If so, which ones? | - added a mention to open research data to help contextualise the question (our focus on open science was already known to participants) | RQ-2: What do humanities researchers think of the word "data" and how do they associate "data" with research materials used in their discipline?<br>RQ-3: What are humanities researchers' attitudes towards open science? |
| 8 | More generally, what would you tell me if I asked you to define the word "data"? | none | RQ-2: What do humanities researchers think of the word "data" and how do they associate "data" with research materials used in their discipline? |

Unlike Thoegersen, we contacted only researchers working within the Department of Classical Philology and Italian Studies [4], not within the faculty at large [5], using the official disciplinary areas published by the Italian National University Council (n.d., pp. 5-7) to select our sample. We contacted via e-mail one researcher for each of the 19 disciplinary areas represented within the department, and they all agreed to talk to us.

We then arranged interviews either in person or via Microsoft Teams, the most widely adopted video-conference platform at the University of Bologna, over 3 weeks, from the 9th of November to the 1st of December 2021. We met 3 researchers in person and 16 online, and all the interviews lasted between 30 and 90 minutes. At the beginning of each, we asked for consent to record for the purpose of transcribing, and to use the transcripts in the study. The transcriptions were anonymised by editing out any content that could be used to identify the interviewees.

## Data analysis

To analyse the interviews, we relied on a grounded theory approach (Glaser and Strauss, 1967) as it can generate rich descriptions, while at the same time rendering interpretive methods more codified and legible to a broad audience (Baumer *et al.*, 2017). The grounded approach is not a codified method but includes several methodologies sharing common epistemological underpinnings (Glaser, 1992).

For qualitative data coding we used QualCoder version 2.8, an open-source application written in python3 that uses a SQlite database to store coding data (Curtain, 2022). The transcriptions, in the original Italian, and the coding are available for reuse on Zenodo (Gualandi *et al.*, 2022).

In this analysis, participants are grouped into 5 main research areas:

- **philology and literary criticism** (12 participants) – includes all the researchers who have texts or documents as their main object(s) of study;
- **language and linguistics** (4) – includes all participants whose research is focused on languages, linguistic phenomena, and language teaching;
- **history of art** (1) – includes the only participant whose research revolves around art history;
- **computer science** (1) – includes the only participant who is a computer scientist;
- **archival studies** (1) – includes the only participant who focuses specifically on archives and archival documents.

Although 4 of the 21 projects described across the interviews could be categorised as digital humanities (DH) projects, as of October 2022 the Italian National University Council does not officially recognise DH as a discipline (n.d., pp. 5-7). Researchers in the DH often encounter challenges around institutional position and career evaluation (Poole 2017), but within Italian Universities they are of necessity assigned to other academic disciplines.

For this reason, we are unable to report if DH practitioners have different views on the concept of data and on open science, or if they manage data in different ways. This is certainly a limitation, considering it is likely that DH scholars are more attuned to the idea of creating and reusing "data" (Poole 2017). One possible solution would have been to ask interviewees whether they self-identified as digital humanists, but unfortunately we did not include this question in the interviews.

In order to answer our first research question (How do we define "data" in the humanities?), we extracted all types of "research materials" mentioned by participants when answering questions 2 and 3a. They often used slightly different words or expressions to describe the same thing, so we normalised the terminology and then grouped materials into 13 different categories (e.g., webinars, conferences and exhibitions were all categorised as events – see also Table II below). We then used these categories to organise the coding so that all the codes pertaining to a specific data type could be grouped together.

As for the second research question (What do humanities researchers think of the word "data" and how do they associate "data" with research materials used in their discipline?), we looked at the answers given to question 7 and 8, using coding to categorise definitions and statements on data as "inclusive", when they described many different research materials as data, and as "restrictive", when they described data as something rather limited and precise. Both kinds of statements often coexisted, so we looked at each interviewee's answer separately, to weigh the different statements and rate their overall opinion on data on a scale from 5 (very inclusive, e.g., "everything is data") to 1 (very restrictive, e.g., "data are numerical"). A similar approach was taken to answer research question 3 (What are their attitudes towards open data and open science?), categorising statements as "mostly positive" or "mostly negative".

Finally, in analysing current data management practices, we used qualitative data coding to let information emerge organically from the interviews. Then, where possible, we grouped the codes into categories that loosely reflected the main topics raised through questions 3b to 6 (e.g., documentation, copyright and privacy, methodologies, etc.).

# Results and discussion

In this section we analyse the results of our interviews. Where relevant, we quote passages from the interviews directly – please bear in mind that all translations from Italian into English are ours.

## Research question 1: How do we define "data" in the arts and humanities?

We extracted the "materials" study participants work with and we grouped them into 13 different data types. This broad definition of data is intended to enrich the understanding of how FAIR data management practices can be implemented in the arts and humanities. At least in principle, all these data types can be both produced and used – or "collected" and "generated", to use Tóth Czifra's terminology (2019, p. 22).

**Table II:** Data types emerged from the interviews

| N | Data types | Produced by (tot. 19) | Used by (tot. 19) |
|---|---|---|---|
| i | publications | 18 | 15 |
| ii | other primary sources (e.g., manuscripts, artworks) | 0 | 18 |
| iii | digital representation of cultural objects (e.g., facsimiles, photos) | 4 | 8 |
| iv | catalogues, databases and other search tools | 2 | 9 |
| v | events (e.g., conferences, exhibitions) | 6 | 0 |
| vi | websites | 4 | 0 |
| vii | software | 2 | 2 |
| viii | documentation | 3 | 0 |
| ix | digital infrastructures (e.g., mobile apps, web platforms) | 3 | 0 |
| x | personal data | 2 | 0 |
| xi | corpora | 2 | 0 |
| xii | standards | 0 | 2 |
| xiii | born-digital artifacts (e.g., tags, associations, texts) | 1 | 1 |

(i) **Publications** are by far the most mentioned material [6]. Across the 21 projects described by our interviewees, the following types of publication have been produced (usually more than one per project):

- critical editions (8 projects)
- journal articles (8)
- monographs or edited books (4)
- essays (3)
- reading editions (2).

This list reflects the fact that the majority of our interviewees are philologists and literary critics. Critical editions and monographs have been kept separate because of their fundamental differences (the edition of a critically reconstructed text in one case, the discussion of a research topic in the other); it is also worth noting that no philologist in our group worked on or with digital scholarly editions.

Further, publications are used both as secondary sources (literature reviews, reference texts, etc.) and primary (especially novels or other literary texts). Apart from publications, (ii) **other primary sources** [7] are:

- ancient manuscripts and early printed books (9 projects)
- modern and contemporary unpublished materials such as drawings, notes and letters (4)
- archival documents (3)
- monuments, artworks and artifacts such as tablets and vases (3).

Their (iii) **digital representations** include facsimiles, photographs and 3D models. Almost half of the study participants use them, although inspecting the objects in person is still considered fundamental. Of the 4 projects that produced a digital representation of some cultural object or event (e.g., videos), only 2 published them for the wider public.

Six interviewees report having organised (v) **events**, often more than one per project, defined as one-off gatherings of people for the purpose of sharing ideas, offering training, or presenting project results to the public [8]:

- conferences (4 projects)
- exhibitions (2)
- webinars (2)
- guided tours (1)
- teacher trainings (1).

We then have at least 3 different types of "electronic paradigm publications" (Edmond 2019, p. 2):

- (vi) **websites** to describe and/or publicise a project, allowing limited user interaction
- (ix) **digital infrastructures** (namely, a mobile app, a 3D model web viewer and a web platform)
- (iv) **catalogues, databases and other search tools,** widely used to study and locate primary sources [9].

All 3 require (vii) **software** to be written and executed, which stands in a category of its own, both in Edmond's classification (2019, p. 2) and in ours.

Finally, we found these additional data types:

- (viii) **documentation**, consisting in deliverables for funders (2 projects) and project notes (1)
- (x) **personal data** [10]
- (xi) **corpora**, a category specific to linguistics research
- (xii) **standards** – languages or other conventions officially recognised as the norm in a certain research field (e.g., for interoperability)
- (xiii) "**born-digital artefacts**", or by-products of users' online interaction with digital infrastructures (e.g., tags, associations, texts) that can be used as input for further research.

These 13 data types represent the "building blocks" of research in the arts and humanities, as emerged from the interviews. In order to successfully apply FAIR principles, these are the type of data to reflect on develop infrastructures for.

While publications are often not considered data (e.g., Harrower *et al.*, 2020; Avanço, Giglia and Gingold, 2021; Prost *et al.*, 2015), arts and humanities scholars we interviewed clearly consider them as the most important type of data they work with. In addition, when we asked study participants to define data, 6 of them explicitly included methodologies in their definition. These are also usually excluded from the definition of data (e.g., Directorate-General EU, 2021; Wilkinson *et al.*, 2016).

## Research question 2: What do researchers think of the word "data" and how do they associate "data" with research materials used in their discipline?

When we asked the 19 participants to our study to think back at the materials listed earlier in the interview, and whether they would consider any of them to be data,

- 14 responded positively – of these, however, 2 expressed unease with the term, and 1 strongly objected to its use,
- 2 researchers answered negatively,
- 3 said they did not know how to respond.

Among those who answered "no" or "I do not know", the concept of data was perceived as linked to the digital, or technological, sphere.

Interestingly, 12 researchers unwittingly used the term "data" before it was first introduced by us in question 7 and 8. Of them, 6 used it exclusively within the term "database" ("banca dati"), 2 referred to personal data ("dati personali"), 2 referred to gathering data as part of their research, while another 2 used it several times and in different contexts.

As mentioned, 2 interviewees expressed unease with the term "data". They stated that it is difficult to define, and that trying inevitably brings up more questions:

> *"I know what [data] are until someone asks me to define them".*

> *"It is a complicated question […] I don't know if I'm answering it or rather asking more questions".*

Another researcher objected strongly to the use of the word, stating that it is generic and useless, in addition to being reductive and linked to the commodification of research. Even so, we found less resistance to the term "data" than we would perhaps have expected.

When asked to define data, most participants seemed to go back and forth between restricting the term to something "circumscribed, objective and well-defined", e.g.,

> *"Data are single information items referring to a problem".*

> *"Data are the raw material of empirical research".*

> *"I have this idea of data as something circumscribed, objective and well-defined, but I don't know if it's correct".*

and taking a more inclusive view, e.g.:

- publications are data:

  > *"a scientific article can be considered data, of course it can be considered as data".*

- methodology and notes are data:

  > *"Methodology is data; and even better if it is innovative compared to those used beforehand, to show it is indeed data, intellectual energy sources that were previously not considered".*

  > *"Reading a passage according to a certain critical framework, as proposed in an essay or an article is data, assuming that is done according to a serious methodology".*

- interpretation and research results are data:

  > *"The results of research can indeed be considered data".*

  > *"I understand data both as data emerged from research and as hypotheses".*

  > *"[…] my judgment on the authenticity of a source can be data".*

- everything is data:

  > *"In my opinion everything is data, everything I produce, handle, reuse – for me they are all data".*

  > *"In my case they are all data, the first look I take at a new manuscript, […] the reasoning on the stemma codicum to understand whether two manuscripts depend on a sub-archetype; everything I obtain from a serious, in-depth investigation is data (but it is a generic term, and as such it is useless)".*

Considering each interviewee's answers in full, we weighed the different statements they made and rated their overall opinion on data on a scale from 5 (very inclusive definition, "everything is data") to 1 (very restrictive, "data are numerical and objective"). The results are shown in table III below.

**Table III:** Definitions of data across the interviews

| Rating | Definition of data | N. of researchers (tot. 19) |
|---|---|---|
| 5 | "everything is data" | 2 |
| 4 | very open, tends to include primary sources, interpretation, methodologies and research results | 4 |
| 3 | open, tends to include either one among primary sources, interpretation, methodologies, research results | 5 |
| 2 | somehow restrictive, but can include either primary sources, critical texts, and/or textual variants | 2 |
| 1 | restrictive: "information items", "raw materials", "quantitative components" | 3 |
| n/a | no reply or could not confidently assign a rating | 3 |

Overall, most researchers gave a somehow inclusive definition (11), while a minority (5) defined data them restrictively as "information items", "raw materials", or "quantitative components". Interestingly, the most restrictive definitions all seem to come from researchers belonging to the field of language and linguistics.

Finally, 2 researchers openly stated they had changed their minds during our conversation. In one case, the interviewee gave a restrictive definition of data as "objective" and "raw"; later, after reflecting on open data policies, they acknowledged that a wider view may also be taken. The other researcher began by considering only research results as data, but then added primary sources to the definition.

## Research question 3: What are arts and humanities researchers' attitudes towards open science?

Based on our inductive methodology, we counted 18 statements that we categorised as "mostly positive", and 10 that we categorised as "mostly negative". Though positive attitudes seem prevalent among participants, some still expressed frustration around:

- the cost of article/book processing charges in open access publishing (1 researcher);
- the new rules around EU-funded projects, particularly the speed at which changes are being implemented and the imposition of open access for monographs, which limits researchers in their choice of a publisher (3);
- the difficulty that less methodologically innovative research projects encounter in accessing funding in the current landscape (3).

Positive statements, on the other hand, concerned:

- a willingness to share research data and a desire to "give back" to society the results of research (4);
- a desire for support from universities and public institutions, especially regarding research infrastructures, eliminating the need to rely on commercial third-party services (2);
- concerns about traditional publishers protecting academic works excessively and in fact preventing the circulation or research results (2);
- a desire to see primary sources digitized and made available for research (2).

Finally, one researcher made an important point:

> "There can be a sense of fatigue for humanists, 'gosh I have to share my data', because to do that I have to follow a standardised schema. […] It is extra work, and I feel we should understand that, if that is the case, all projects need a bit more workforce and even people with specific expertise. We have very small projects."

These points echo those made several times across the literature (Tóth-Czifra, 2019; Edmond, 2019; Harrower *et al.*, 2020): that is necessary to work on current evaluation practices to make data curation worthwhile, and that institutions need to provide support to humanities researchers in order to realise the FAIRification of research data (GO FAIR, 2022).

## Research question 4: What are arts and humanities researchers' current data management practices?

### *The widespread use of digital resources and technologies*

Five interviewees stated that the digitisation of manuscripts and archive inventories radically changed their way of working with primary sources. In some cases, digital copies allow access to otherwise inaccessible resources (e.g., due to geographical distance, geopolitical issues), in others they make it possible to better preserve the original documents, when they are too frail to be accessed and consulted with ease.

However, philologists or literary critics in our cohort tend not to include any images of the source documents within the final research output, especially when this is a critical edition. On the contrary, the one art historian in our sample stated that digital images of artworks are fundamental for the discipline and should be available without copyright restrictions:

> *"The problem of image copyright is a catastrophe because the cost of one single image can derail any budget. […] A photograph can cost 200 €, and you can imagine [what happens] if one needs seven or eight of them".*

Many participants commented on the level of digitisation of our cultural heritage, with a particular focus on the work being done by libraries and archives. If one researcher complimented the digitisation campaigns undertaken by several important Italian libraries, many more defined digitisation efforts as slow and insufficient, e.g.:

> *"Digitisation of archival documents is really behind […]".*

> *"Only a small fraction of these materials is digitised, and online catalogues are also lacking".*

> *"[…] human resources are missing, competences are missing. But for us doing this kind of research it would be fundamental, it would be a turning point."*

More than one interviewee commented that technology has radically changed their work and another one talked about using technology to push methodological boundaries. Across our sample, digital research tools and digital representations of cultural objects are extensively used. A small minority of researchers (2/19), however, expressed resistance to the replacement of physical printed books with digital publications stating, e.g.:

> *"The digital part is very important and can be really useful, even to structure the edition and show things, but then it is necessary to keep printing and reading [on paper], not only for the pleasure of it, but because some in-depth analysis can only be done on that kind of support".*

## *A wealth of methodologies*

When asked whether their work is based on a precise methodology, recognised across the field, 11 of the 12 **philologists and literary critics** in our sample immediately replied "yes". E. g.:

> *"Absolutely, the discipline has its shared standards, even very formal standards because classical philology is really interested in the formal aspects".*

> *"We too have a method, some forget about it, but a 'bad' article is an article that has not followed the method, that advances hypotheses that cannot be demonstrated, that has logical loopholes".*

We found a multiplicity of methodologies in this area, which are reportedly mixed and matched according to the literary work under study. Just to give a couple of examples, philology of copy makes ample use of Lachmann's method and of the *stemma codicum* to reconstruct the original text, while authorial philology is mostly based on a single witness which is rarely, if ever, edited.

As far as the description and formalisation of these methodologies is concerned, some participants indicated they happen in university manuals, as well as in preceding works that have become "milestones"; others described methodologies as "deposited through practice".

Also, critical editions are described as deeply formalized research outputs. They have a precise structure, although there are some "margins to experiment" since the primary sources being edited are extremely varied:

> *"Within a critical edition I expect to find the text presented in a certain way, an apparatus compiled in a specific way, and I know the type of information I'm going to find there".*

> *"[…] of course, there are parameters to evaluate if a critical edition is scientifically sound or not".*

One researcher highlighted what, in their view, is a fundamental difference between philology and criticism:

> *"I am a philologist, so I'm the closest to science. Criticism is further away, there clearly is more freedom of choice […]. Those who work in philology are closer to scientists and move in a less extemporaneous way, we have templates".*

Interestingly, the only researcher in this disciplinary area who stated that they do not follow a method is indeed a literary critic, who questioned the very idea of a method:

> *"I do not believe in a pre-constituted method, I come from a school that did not believe in it […]. So, I think that even the sciences do not have a pre-constituted method, that breakthroughs happen by infringing on a method".*

**Language and linguistics scholars** (4) also described different methodologies, corresponding to different approaches,

> *"In psycholinguistics and neurolinguistics […] the protocols are close to those of the experimental sciences in the medical field […], historical linguistics has more traditional methods, only partially codified."*

together with a combination of qualitative and quantitative methods. One of them stated that best practices are often more important than theoretical considerations, due to the applied nature of the discipline.

Here, too, the formalisation of methodologies is described as scarce:

> *"It is still left to the good will of individual scholars".*

> *"I wouldn't say there is a standard […] but there certainly are recurring elements".*

Linguistic data is rarely (but increasingly) shared together with aggregated results, and this hesitation is linked to privacy concerns and to the extra work involved:

> *"The materials, the recordings, are interesting but I cannot ask [participants] to sign a blanket consent […] because I would risk decimating the, already few, people who are enthusiast to participate […]; the dissemination will thus be limited".*

The observations made by the one **art historian** in our group cannot of course be considered representative but are nonetheless interesting. Here, too, different approaches are described, while audio-visual materials are naturally considered very important. In addition, the researcher talked about "visual philology" providing a shared methodological basis:

> *"I consider fundamental this aspect that I would call our philology, the evaluation […], it is a real visual analysis […]. I believe it is the fundamental basis of our discipline: we all share the same templates to recognise different styles, different languages […]. Because 'visual philology' is real philology, otherwise it is easy to make mistakes".*

The one researcher working in **archival studies** stated that shared methodologies exist but are not clearly formalised. One of the main reasons is that all archives are unique, having in some cases been built over several centuries.

Finally, the only **computer scientist** in the group stated that there are of course clear methodologies and paradigms, which, again, are mixed and matched according to each project's needs. Here, methodological choices are routinely documented in software documentation and deliverables, helping the reuse and modification of pre-existing code.

## *The importance of collaboration and sharing*

Collaboration emerges as pivotal for our group of researchers, contrary to what is reported in similar studies (Akers and Doty, 2013, p. 10; Thoegersen, 2018, p. 495). It takes place through email and telephone communications, cloud storage and file-sharing services (often Google Drive), or other specialised software (e.g., for corpus creation).

In many cases (9/19), it simply entails consulting with other researchers, within the same disciplinary field, or not. This type of collaboration, informal and not recognised in academia beyond the occasional acknowledgment within the final publication, is described by one of our interviewees as "not 'real' sharing" because it comes down to asking for feedback and suggestions rather than opening up to real collaboration:

> *"[…] sharing is something else, is letting somebody else into your own research project".*

However, 12 researchers in our sample, across all disciplinary areas, stated that they work, or have worked, in team with others (mostly colleagues, sometimes PhD and master's students). A couple of them mentioned that they "like working alone" or that "philologists often work alone", but they still described teamwork as "extremely useful and productive". Only one researcher, a literary critic, stood out from the rest of the sample by stating:

> *"I do not share my work materials with anybody, and my colleagues do the same, perhaps we are a bit too archaic. There is no sharing, but rather a 'top secret' kind of approach. [...] I also have colleagues who fear their ideas might be stolen [...], sharing is neither habitual nor common for us."*

The fear that ideas might be stolen before publication was mentioned by 5 interviewees in total, 2 of which explicitly linked it to research evaluation. E.g.:

> *"It is better to keep the unpublished archival original to yourself, otherwise somebody can publish it before you do".*

> *"The material is shared with very few people and only partially, not in its entirety, to protect the originality of the research and its novelty, which has to be made public only when the research is finished and complete".*

> *"I would share research results with anybody but [...] as long as hiring, career progression and evaluation within universities [...] keep happening with the current parameters I am forced to use copyright protection and keep results inaccessible".*

> *"I like putting anything in common, but we face an issue of intellectual property since we mostly publish single-authorship articles, and our career is based on publications".*

## *Specialised content, often inaccessible*

Unsurprisingly, data produced by our interviewees are mostly intended for a specialised audience. Some are intended for a broader public (e.g., reading editions, websites, exhibitions and tours), while others are created with a very specific audience in mind, such as the project team (e.g., deliverables, project notes, personal data). In most of the cases brough to our attention by the study participants, this kind of restrictions depend on privacy concerns.

Table IV provides a summary assessment on the accessibility of the different kind of data, as derived from the interviews.

**Table IV:** Summary assessment on data accessibility

| Research output | Assessment on accessibility |
|---|---|
| software | freely available online |
| standards | |
| websites | |
| born-digital artifacts (e.g., tags, associations) | usually freely available online but access may be restricted |
| catalogues, databases and other search tools | |
| digital infrastructures (e.g., mobile apps, web platforms) | |
| documentation | |
| digital representation of cultural objects (e.g., facsimiles, photos) | if published at all, they are usually freely available online; otherwise, permission to publish must be sought and can be expensive |
| primary sources different from publications (e.g., manuscripts, artworks) | mostly held in conservation institutes; depending on rarity and state of conservation, they can be accessible to the public, or only to scholars under supervision |
| events (e.g., conferences, exhibitions) | may require registration and relevant specialisation (unfortunately, we did not gather any information on this point during the interviews) |
| publications | mostly published in closed access |
| personal data | for internal use only |
| corpora | |

If we look in more detail at publications by our participants, we see that:

- the 8 critical editions and 2 reading editions mentioned in the interviews are published on paper, in closed access, and with a traditional publisher,
- 1 out of 4 edited books has been published in open access, for express wish of its authors,
- 2 out of 8 projects producing journal articles made them available in open access, according to funding requirements.

Regarding the accessibility of linguistic corpora, please see the next section.

*Copyright and privacy issues*

We have already mentioned how image copyright was flagged as problematic, while a couple of researchers also lamented the different copyright laws across different countries, which complicate international collaborations. Another participant reported doing extra work to avoid dealing with copyright issues:

> *"I have added translations to make my work more accessible. I have translated everything from scratch because I did not want to face the issues of using somebody else's translation".*

All corpora produced by the language and linguistics scholars we interviewed are associated with and/or contain personal (and sometimes sensitive) information and thus are not made available to anyone outside of the project team. Our interviewees recognised that value of making these corpora available to the research community through anonymisation, and stated that the discipline is moving in that direction.

*Researchers keep everything (private)*

All researchers but one stated that they save most data used and produced during a research project because they are potentially useful both in teaching and for further research. E.g.:

> *"The projects I have followed […] have always been linked to one another; you do not really finish and throw everything away, there is always an element of going back to things that it may be useful to retrieve".*
>
> *"I very often use materials from previous or ongoing research projects in my teaching".*

However, only 2 participants mentioned using an (open) repository, in both cases Zenodo, while they all reported storing digital data on their own devices, especially external hard drives, or on commercial cloud storage services (Google Drive and Dropbox were the most commonly mentioned).

### *The absence of documentation and standards*

When asked whether the materials mentioned in the interview had any accompanying documentation, 15 researchers answered negatively or simply ignored this part of the question, while only 4 responded affirmatively. They mentioned:

- software documentation (1),
- deliverables for funders (2),
- documentation shared among the project participants (1).

The situation is not dissimilar when looking at standards. At the time of the interview, we chose not to define the word, letting researchers free to interpret it as they pleased. Most participants (11) understood it in a broad sense and proceeded to describe how their discipline does indeed have "standard" methodologies, recognised across the field (for a discussion on methodologies, see above). Three interviewees simply responded there are no standards in their field, and 3 more mentioned archival and museum standards in relation to the primary sources of their research.

However, 2 researchers mentioned:

- standards such as JSON, RDF, XML and OWL (1),
- cataloguing and technical standards to ensure interoperability between databases (1).

Perhaps unsurprisingly, documentation and standards are thus familiar only to a small subset of researchers, who do not belong to the philology and literary criticism group.

# Conclusions

Stretching the definition of data to include publications, workflows or indeed anything researchers work with may seem dangerously vague and ultimately useless.

But if we talk about FAIRifying research data, then this concept must include all materials used and produced in arts and humanities research. Rather than suggesting a different terminology, we think it is more useful to reflect on how we can define data in productive way since "Good definitions […] underpin our paradigms, theories and frameworks; support our research and practices and further our understanding of the phenomena they define" (Boyd 2022, p. 1339).

While we are not suggesting any practical solutions for FAIR data management in the humanities, the extremely broad definition of data that we propose needs to be understood within the context of the FAIR principles and of the requirements around research data management increasingly coming from funding bodies. Potential implications of our findings on data management practices will be discussed in future publications, with a focus on the differences between the individual communities within the humanities at large.

In this study, we have looked closely at how researchers work day-to-day, "going as deep as possible in research practices" and linking the definition of data "to the workflow of knowledge creation" (CO-OPERAS and Giglia, 2020, p. 1). We have done so by interviewing researchers about their work, their research data are their current data management practices. Our experience is of course limited to a small sub-set of researchers and research areas within the arts and humanities – it should be replicated on a larger scale, and include other disciplines, in order to paint a reliable picture of the humanities. However, to transition to open science, it is extremely important to engage academic communities (Armeni *et al.*, 2021), however small, and a project of this kind can indeed serve this purpose well.

We were able to recognise at least 13 types of data that are collected and/or generated by the researchers we interviewed. Publications emerged as the most important one, hinting at how open access to scientific publications and open data policies are strictly interconnected in the humanities.

Wondering whether "data" is still a dirty word in the humanities, Tóth-Czifra, pointed out that "[s]urveys from the past 5 years reveal a great deal of uncertainty in arts and humanities researchers' conceptions of data and its applicability to their own work" (2019, p. 15). Indeed, only a slight majority of respondents at Minnesota Twin City considered the term "data" applicable to their research (Hofelich Mohr et al., 2015, p. 52) and Thoegersen found "a general uncertainty among the participants as to what falls into the scope of data" (2018, p. 501). While our 19 interviewees also showed a good deal of uncertainty, they ultimately did not find the term "data" problematic: 14 of them considered it applicable to their research and only 3 expressed unease with or resistance to the term.

To them, data does not seem to be a "dirty word" and talking about "FAIR data" and "FAIR data management" in the context of humanities research may not be counterproductive after all. Rather, we need to make sure that the term "data" is intended by everyone involved in a broad sense, one that is inclusive of all modes of research. This is what most of our interviewees (11) did, and seems to us the only way to "help to realise the promises of FAIR guidelines in an arts and humanities context" (Tóth-Czifra 2019, p. 24).

Attitudes towards open science tend to be positive across our pool of researchers. However, as stated by a handful of interviewees and as emerged clearly from the literature (CO-OPERAS *et al.*, 2020; Edmond, 2019; Tóth-Czifra, 2019), it is necessary to work on current evaluation practices and support humanities researchers in order to make data curation worthwhile.

Regarding other aspects of data management, our results differ from previous studies (Akers and Doty, 2013; Thoegersen, 2018) as far as the willingness to share research data with colleagues is concerned: we found collaboration to be extremely important for our interviewees, both in the form of informal sharing and of teamwork. A possible explanation might be found in the influence of digital humanities projects, which are intrinsically more collaborative as they "require human and material resources spanning disciplinary and institutional boundaries" (Senseney, Dickson Koehl and Nay, 2019). However, as explained in the methods section, we were unable to analyse digital humanities as a separate disciplinary area.

Finally, when we asked researchers whether their work is based on a precise and well recognised methodology, the vast majority replied positively. Although we are a long way away from workflows that are "explicit […] documented and, therefore verifiable" (Edmond, 2019, p. 15), our analysis suggests that researchers might be ready to reflect on how to "characterize humanities workflows" and to "identify how such characterizations can be made useful" (Fenlon, 2019, p. 512).

---

1 https://allea.org/
2 https://www.go-fair.org/
3 https://operas.hypotheses.org/operas-p
4 https://ficlit.unibo.it/it
5 The School of Arts, Humanities, and Cultural Heritage at the University of Bologna includes 3 other departments: Cultural Heritage, Philosophy and Communication, and History and Cultures.
6 These correspond to Edmond's "print paradigm publications" (2019, p. 2).
7 Corresponding to Edmond's "single or collected/curated primary sources" (2019, p. 2).
8 They correspond to what Edmond (2019, pp. 2-3) calls "ephemera".
9 Although these could in principle be physical objects, they are overwhelmingly digital in our experience. Nine participants report using them to locate a primary source, find information about it, and sometimes access its digital representation. Examples include bibliographic and lexical databases, lexicons and dictionaries, online catalogues; some contain digitised documents (Internet Archive, Google Books). We realise this is an extremely broad category and should be divided up further.
10 For a definition see: https://ec.europa.eu/info/law/law-topic/data-protection/reform/what-personal-data_en